\documentclass{desyproc}
\usepackage{hyperref}
\usepackage[normalem]{ulem}
\usepackage{color}
\usepackage{cite}

\begin{document}
\title{\vspace{-2.05cm}
\hfill{\small{{DESY 13-184, MPP-2013-297}}}\\[1.27cm]
Dark Matter -- a light move}

\author{{\slshape  Javier Redondo$^{1,2}$,  Babette D\"obrich$^{3}$ }\\[1ex]
$^1$ Arnold Sommerfeld Center, 
Ludwig-Maximilians-Universit\"at,  M\"unchen, Germany\\
$^2$ Max-Planck-Institut f\"ur Physik, M\"unchen, Germany 
$^3$ Deutsches Elektronen-Synchrotron (DESY), Hamburg, Germany \\
}

\contribID{redondo\_javier}

\desyproc{DESY-PROC-2013-XX}
\acronym{Patras 2013} 
\doi  

\maketitle

\begin{abstract}
This proceedings contribution reports from the workshop Dark Matter - a light move, held at DESY in Hamburg in June 2013. 
Dark Matter particle candidates span a huge parameter range. In particular, well motivated candidates exist also in the sub-eV mass region, for example the axion. Whilst a plethora of searches for rather heavy Dark Matter particles exists, there are only very few experiments aimed at direct detection of
sub-eV Dark Matter to this date. The aim of our workshop was to discuss if and how this could be changed in the near future.
\end{abstract}

\section{Introduction {and light Dark Matter theory}}

The standard model of particle physics (SM) has passed its last test with the recent discovery of the Higgs boson, 
but it is still incomplete as a full theory of nature. The most urgent claim of physics beyond the SM  
is the existence of non-baryonic dark matter (DM) in the Universe, most likely new particles whose nature we have failed to unveil so far. 
Extensions of the SM originally proposed to convert it in a more natural choice among conceivable theories,
suggest a few DM candidates: weakly-interacting massive particles (WIMPs) or the axion, among the best motivated. 
To date, a major part of the community hopes for direct detection of DM in experiments aiming to record the 
recoil energy of nuclei after scattering with DM particles. 
This strategy is optimal for WIMPs with masses $\gtrsim 100$ GeV but leaves out very light DM candidates 
such as the axion, searched for by just one experiment (of a very different nature): ADMX. 
While there are many efforts to discover WIMPs in the near future, 
ADMX can only cover a part of all the viable axion DM parameter space. 
Moreover,  axions are just an example of a broad class of particles, 
weakly-interacting slim particles (WISPs), that share most of their phenomenology~\cite{Jaeckel:2010ni}. In particular, WISPs are excellent DM candidates~\cite{Arias:2012az}. New experiments have been proposed to search for WISPy DM~\cite{Baker:2011na,Graham:2013gfa,Budker:2013hfa,Horns:2012jf,Jaeckel:2013eha,Sikivie:2013laa}, sometimes sensitive to more than one type of DM candidate at a time, e.g. axion-like particles and hidden photons. 
These novel experiments can explore decades of pristine
DM parameter space and pave the road to boost the sensitivity of ADMX-like experiments 
looking for axions in unexplored  parameter regions. 

\begin{wrapfigure}{r}{0.5\textwidth}
\centering
\includegraphics[width=0.5\textwidth]{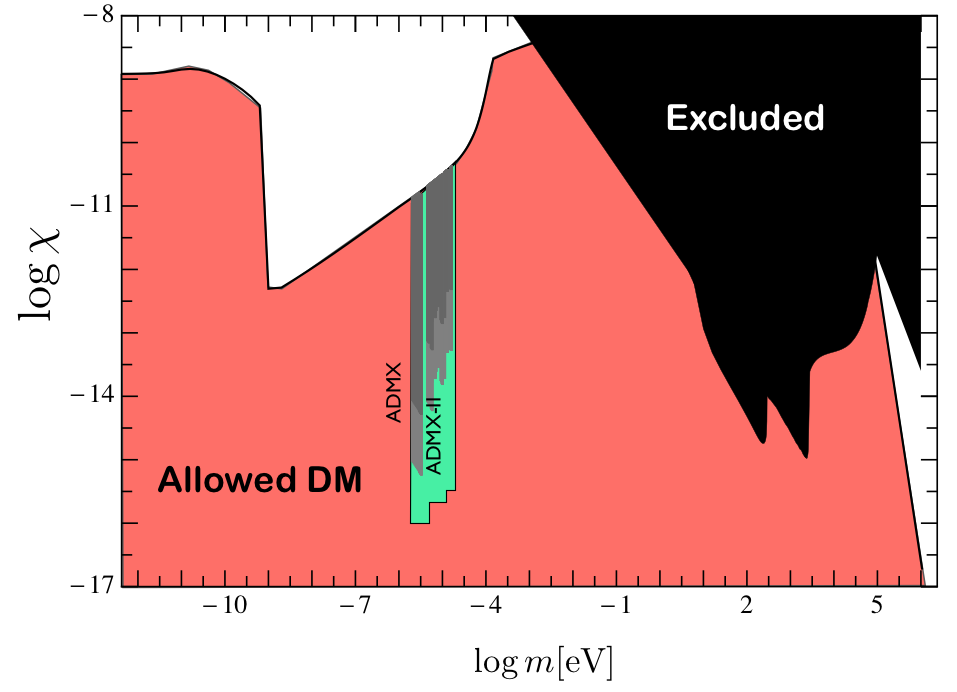}
\caption{\small Parameter space for HPs where they can account for the cold DM of the Universe. 
Also shown are excluded regions and the forecasts of ADMX and IAXO.}\label{Fig:MV}
\label{parspace2}
\end{wrapfigure}

The workshop `Dark Matter -- a light move' took place at DESY Hamburg at June 17th-18th~\cite{address} with the 
purpose of fostering new direct detection experiments looking for axions and other WISPs.
The gathering was a blend of theorists and experimentalists expected to team up in selecting the most promising regions of parameter space and realistic set-ups that can cover them.
The expertise of the audience was chosen to focus on exploiting the couplings of WISPs to photons,
which present several advantages.
At the theoretical level, WISPs appear in well motivated field and string-theory extensions of 
the SM~\cite{Ringwald:2012hr}. 
Axions and axion-like particles (ALPs) appear as pseudo-Nambu-Goldstone bosons of 
spontaneously broken global symmetries at high energy scales and as imaginary parts of moduli fields in 
string-theoretic theories
or in general in theories where the sizes of gauge couplings are set by the vacuum expectation value of new fields. Their coupling to photons is of the well-known kind
\begin{equation}
{\cal L}_\phi = \frac{1}{4} \ g_{\phi \gamma}   \phi \ F_{\mu\nu}\tilde{F}^{\mu\nu}, 
\end{equation} 
where $\phi$ is the axion (ALP), $g_{\phi \gamma}\sim \alpha/(2 \pi f_\phi)$, and  $f_\phi$ a symmetry breaking energy scale and $F_{\mu\nu}$ the electromagnetic tensor.
The mass of axions (ALPs) depends upon terms that explicitly break the global symmetries. 
For the axion we have $m_a=6\rm\ meV(10^9\ {\rm GeV}/f_a)$ with $f_a$ the symmetry
breaking scale of the axion, while for ALPs 
such a relation is much more model dependent. 
Hidden photons (HPs) appear also in field and string-theoretic extensions \cite{Jaeckel:2013ija}. Their main interaction
with photons is through kinetic mixing
\begin{equation}
{\cal L}_\chi = -\frac{\chi}{2}F_{\mu\nu}B^{\mu\nu},  
\end{equation} 
where $B^{\mu\nu}$ the HP field-strength and $\chi$ is the kinetic mixing, 
with typical values in the $10^{-12}\sim10^{-3} $ range. WISPy cold dark matter 
can be produced in the early Universe by a variety of mechanisms. The most relevant 
are the misalignment mechanism and the decay of topological defects. 
The regions of mass-coupling  parameter space where the full DM can be accounted 
for the different cases were summarized in the workshop and is 
shown in figs.~\ref{parspace} and~\ref{parspace2}. The most important 
constraints on WISPy DM come from: non-observation of WISP DM decay, absorption
of WISP DM in the early universe plasma and indirect effects on stellar cooling (non-DM WISP emission), 
from refs.~\cite{Cadamuro:2011fd,Jaeckel:2013ija} which we have shadowed in black in the figures.  
A DM WISP background imprints generically isocurvature anisotropies (generated during inflation) in the cosmic microwave background which have not been observed, imposing strong constraints in WISP DM models and the 
parameters of inflation~\cite{Folkerts:2013tua}. 
Finally, the WISPy DM paradigm could be particularly well tested 
because WISPs can form a Bose-Einstein condensate~\cite{Erken:2011dz}
and thus lead to the formation of peculiar caustics in 
galaxies~\cite{Sikivie:2010bq}, which could have already 
been observed~\cite{Banik:2013rxa}. 

\section{Detecting WISPy DM}

DM axions (ALPs) mix with photons in a background magnetic field (strength $B$) with an angle
\begin{equation}
\chi_\phi = \frac{g_{\phi \gamma} B }{m_\phi}  \quad ({\rm generic\ ALP}) \quad;\quad 
\chi_a = \frac{g_{a \gamma} B }{m_a} \sim 10^{-15}\left(\frac{B}{10\ \rm T}\right)C_{a\gamma} \quad ({\rm axions}) ; 
\end{equation} 
with $C_{a\gamma}\sim \mathcal{O}(1)$, while for HPs the angle is simply $\chi$. 
The local density of DM, $0.3$ GeV/cm$^3$ implies an electric field $|E| \sim 2.3\times \chi$ kV/m 
and it holds for the DM mass $m$, and its frequency, $f= 240 {\rm MHz}\times (m/{\rm \mu eV})$. The $E$-field 
carried by the DM WISP drives reflected waves from mirrors, emitted perpendicularly to the surface to a large degree due to the spatial coherence of DM waves~\cite{Horns:2012jf}.  
The simplest experiment to concentrate this radiation is a spherical dish with a detector at 
its center, getting a radiated power per dish 
area $\sim \chi^2|E|^2= 1.4\times 10^4\chi^2$ W and $\sim 10^{-27}(B/{\rm 10\  T})^2C_{a\gamma}^2$ W 
for axions (note that for axions and HPs, the radiated power is independent of the mass). 

\begin{figure}[h]
\includegraphics[width=.50\textwidth,height=7cm]{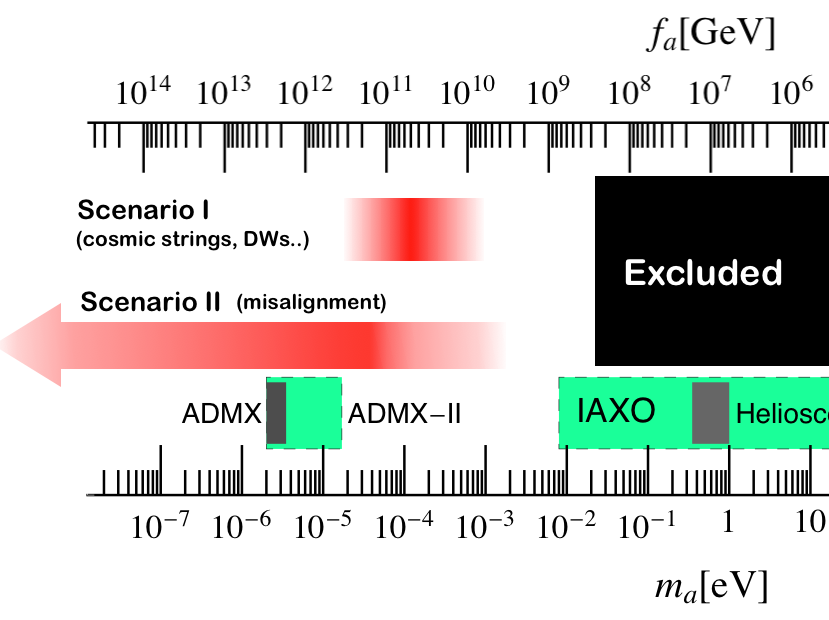}  
\includegraphics[width=0.50\textwidth,height=7cm]{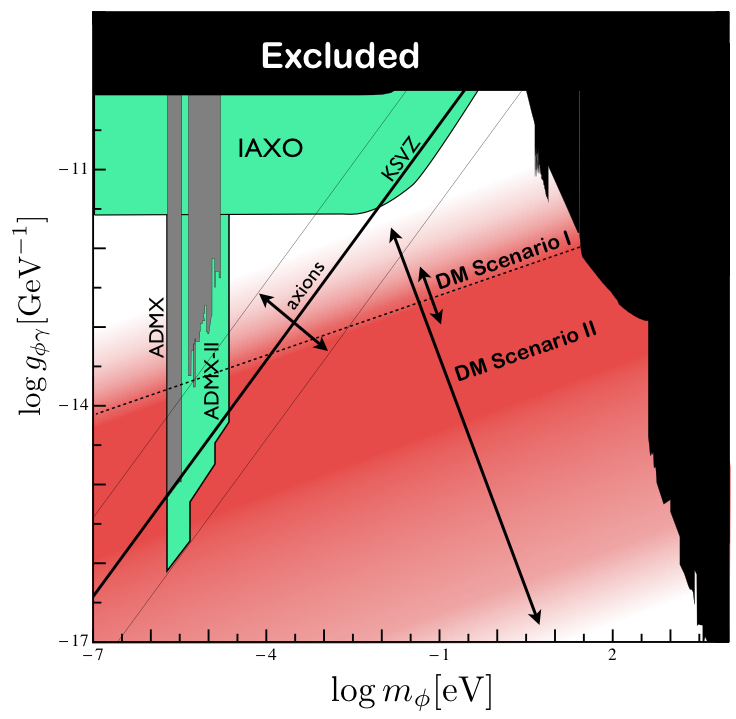}
\caption{\small Parameter space for axions (left), ALPs (right) where they can account for the cold DM of 
the Universe together with excluded regions and the forecasts of ADMX and IAXO~\cite{iaxo}.}
\label{parspace}
\end{figure}

The emitted power can be amplified in a resonant cavity up to a factor of $10^6$ (the
inverse width of the DM energy distribution) but since the WISP mass is unknown, one is forced to scan over frequencies in search for the tiny signal~\cite{Bradley:2003kg}. 
ADMX~\cite{Asztalos:2009yp} is nowadays the only haloscope~\cite{Sikivie:1983ip} of this kind.  
It employs a cylindrical cavity tunable with internal rods inside a 8T solenoid. 
Its dimensions (1 m long, 0.5 m diameter) set its lowest resonant frequency 
$0.48$ GHz ($m=2\ \mu eV)$. ADMX has already taken data in the $2-3.6 \mu$eV 
mass range with a low-noise SQUID amplifier reaching a system temperature $T_S\sim 3$K.
Cooling with 
a dilution refrigerator is planed to achieve $T_S\sim 200 \mu$K, which will be used 
to scan over masses with unprecedented sensitivity (ADMX-II). The first two cavity 
harmonics will be scanned in parallel, masses $2-9 \mu$eV. 
R\&D is taking place at Yale U. to produce haloscopes sensitive at higher masses 
(to cover the Scenario-I, see fig.~\ref{parspace} left hand panel) using superconducting hybrid cavities and Josephson parametric amplifiers working below the quantum limit (ADMX-HF).
A first setup will take data in the 4-8 GHz range ($17-33 \mu$eV) and new designs are being pushed with the ambition to cover up to 20 GHz. 
These prospects are shown in figs.~\ref{parspace} \&  \ref{parspace2} as green regions, IAXO is elaborated in the next section.

\section{New experiments}  

Listening at the workshop to the future plans of ADMX, it was quickly realized
that rather than competing with ADMX in the 2-10 $\mu$eV mass 
range search for axions, the community would benefit of complementing
it at different mass ranges and developing new approaches such
as the dish-antenna experiment. 
First efforts in the first direction have started with WISPDMX \cite{Horns:2013ira},
a cavity experiment~\cite{Bradley:2003kg} based on a HERA proton accelerator cavity. 
Two modes at 208 and 437.3 MHz ($\sim 0.9$ and $1.8\ \mu$eV mass) will be recorded simultaneously 
with a tunning range $\sim 2\%$ provided by 5 plungers. 
A HP run is foreseen next year and magnet options for an axion(ALP) search at DESY or CERN were discussed. 
At the latter facility, the considerable experience 
in microwave light-shining through a wall setups~\cite{Betz:2013iib} 
could be used to boost the sensitivity and tuning range. 

An exciting opportunity to look for WISP DM could also arise with IAXO, see~\cite{iaxo}.
IAXO's main purpose is detecting the solar axion flux, but the required toroidal magnet could host one or several long-cavity experiments or dish-antenna searches. The intense $B$-field (up to 5 T), the gigantic volume (8 cylindrical bores 20 m long and 0.6 m diameter) and the already implemented cryogenics are very desirable for DM searches. 
A long rectangular cavity (up to 20 m long and 0.42 m wide) can be fit into one bore and scan masses above $1.5\ \mu$eV in the spirit of ref.~\cite{Baker:2011na}. 
Tuning strategies are under discussion. 
As a pathfinder experiment, it was proposed to install a similar but smaller  
wave-guide in the HERA dipole used by the ALPS-I experiment for its 
light-shining-through walls experiment\footnote{For the
HERA magnet at the magnet test-bench,
the cavity option would profit from
the fact that the beam-pipe was straightened for 
the future second ALPS setup 
\cite{Bahre:2013ywa}.}, aiming at
higher masses ($\sim 20\ \mu$eV). Such an experiment is under discussion. 
Among the workshop participants,
expertise in cavity construction, tuning,
operation and 
signal detection was quickly put forward.

A good amendment to the cavity option could be a
broadband dish search \cite{Horns:2012jf},
see also the next section. A good deal of the workshop addressed the possible
detector options for this undertaking.

\section{A little workshop aftermath}


In addition to WISPDMX at DESY, a small collaboration has formed from workshop participants to realize the first dish-antenna 
experiment searching for HP (and eventually ALP) DM. 
It will be a proof-of principle experiment: quick and cheap but having still an acceptable reach in the WISP Dark Matter
parameter space. For this goal, we have successfully applied for
additional support from the Helmholtz Alliance for Astroparticle Physics.

The most promising range for such a setup is the mm-wavelength regime (meV mass). 
The unexplored allowed parameter range for HP DM starts at rather large couplings $\chi\sim 10^{-9}$  
whilst for the ALP DM, available detector sensitivities are promising to tackle viable parameter range. 
However, good detectors in the millimeter range are costly and less sensitive 
with respect to photon numbers
than in the optical regime. Due to this fact, and due to the fact that 
the collaborators of this project are highly experienced in optical techniques,
the  dish/mirror Dark Matter search will be performed in the optical regime.
With available equipment, eV photons down to $\chi \sim 10^{-12}$ should be accessible. 


For the search for ALPs, the mirror setup will be  embedded in a magnetic field.
The unprobed parameter range for ALPs in the optical regime is even harder
to tackle than for HPs. However, running the envisaged setup also
in a magnet environment (superconducting solenoids with large bore are in principle
available at DESY) will yield insight in the experimental complications arising
from this demand and eventually pave the way for a dedicated ALP DM search
in the mm-wavelength regime.

In summary, there are many experimental options 
available to check the sub-eV range for Dark Matter -- we look forward 
to see more such options realized and to eventual findings that could solve the Dark Matter puzzle.

\section{Acknowledgments}

The authors acknowledge the Helmholtz Alliance for Astroparticle Physics 
for funding the light-move workshop, Igor G.~Irastorza and Axel Lindner
for their initiative for it, and all participants
and co-organizers
for contributing their ideas. We also thank the organizers of Patras 2013 for a stimulating workshop.
JR acknowledges support from the Alexander von Humboldt Foundation and 
 by the European Union through the Initial Training Network ``Invisibles''.


\begin{footnotesize}

\end{footnotesize}


\end{document}